# Room-temperature THz photon detection via nonlinear upconversion with 2% full-system efficiency

Aswin Vishnuradhan,[1*] Wei Cui,[1] Hesam Heydarian,[1] Eeswar Kumar Yalavarthi,[1] Nicolas Couture,[1] Alain Villeneuve,[2] Angela Gamouras,[1,3] and Jean-Michel Ménard[1,3,*]

[1]Department of Physics, University of Ottawa, Ottawa, Ontario K1N 6N5, Canada
[2]Consultants Optav Inc, Mont-Royal, Québec, Canada.
[3]National Research Council Canada, 1200 Montreal Road, Ottawa, Ontario K1A 0R6, Canada

**ABSTRACT**. Sensitive detection of terahertz (THz) radiation is fundamental to progress in spectroscopy, advanced wireless communication, and the realization of emerging quantum technologies. However, the intrinsically low photon energies in the THz range combined with thermal background radiation tend to constrain detector performance when operating at ambient temperatures. Here, we demonstrate efficient room-temperature THz detection based on nonlinear upconversion in the organic crystal *N*-benzyl-2-methyl-4-nitroaniline (BNA) to resolve frequencies from 1 to 7.5 THz. The system encompassing spectral filters and a single-photon counter achieves an overall detection efficiency of 2% for sum-frequency generated photons. This enables the detection of a train of 50 000 terahertz pulses carrying, on average, fewer than 0.04 photons per pulse, with a signal-to-noise ratio of unity. At a higher flux, when ~60 photons per pulse impinge on the BNA crystal, the per-pulse detection probability reaches 50%. After accounting for loss mechanisms in the setup, the nonlinear THz-to-near-infrared conversion efficiency in BNA exceeds 75%. These results demonstrate the feasibility of quantum experiments relying on single-photon-level THz detection via upconversion in nonlinear crystals in ambient conditions.

## I. INTRODUCTION.

Ultra-sensitive optical detectors, most notably single-photon counters, have driven major advances in quantum communication, quantum optics, astronomy, and remote sensing [1–3]. While conventional semiconductor detectors offer high sensitivity and low noise at near-infrared (NIR) frequencies and above, room-temperature direct detection of terahertz (THz) radiation is challenging because the THz photon energy $h\nu$ becomes comparable to the thermal energy $k_B T$ [4]. Sensitive THz detection plays a vital role in applications such as non-destructive material characterization [5], biomedical imaging [6], high-capacity wireless communication [7], and emerging THz quantum technologies [8]. Although superconducting [9] and quantum-dot-based detectors [10] can achieve single-photon sensitivity at THz frequencies, their broader use is restricted by the requirement for cryogenic cooling.

Parametric upconversion of THz radiation to the NIR provides a thermal background-free method for sensitive THz detection [11–14]. This approach has been implemented in a range of nonlinear media, including inorganic crystals [11,13–16], organic crystals [12,17,18], and Rydberg atomic vapors [19]. Beyond classical sensing, parametric processes have also enabled THz quantum sensing through interferometric schemes based on entangled THz–visible photon pairs, in which the THz photons probe the sample while the correlated visible photons are detected [20]. In Rydberg atomic-vapor systems, THz-to-optical upconversion via multi-wave mixing can enable room-temperature, single-photon-level THz detection. However, the reported demonstration is restricted to the sub-THz regime and a narrowband operation, requiring multiple continuous-wave (CW) lasers and an auxiliary microwave field [19]. In contrast, solid-state crystals can support broadband, pulsed upconversion in a comparatively simple optical setup. THz-to-NIR upconversion in $LiNbO_3$ nonlinear crystals has demonstrated minimum detectable THz energies on the order of 130 zJ at 1 THz by employing multi-stage parametric schemes [13]. Further improvements in sensitivity have been achieved in GaP by combining THz-to-NIR upconversion with single-photon counting, enabling zeptojoule-level detection with 1 s signal averaging, as well as the resolution of individual THz pulses with attojoule-level energies at 2 THz [14]. The pursuit of reaching broader operational bandwidth and even higher sensitivity motivates the exploration of nonlinear media beyond GaP.

Organic nonlinear crystals offer an attractive alternative because they combine large second-order susceptibilities, low THz absorption, and phase-matching conditions that support broadband operation. Demonstrations based on DAST [12,21], DSTMS [17], and OH1 [18,22] have shown that organic materials can be used to achieve efficient THz upconversion in the NIR region. However, many of these techniques rely on frequency-tunable, narrowband difference-frequency generation (DFG)-based THz sources rather than broadband pulsed emitters. In addition, these crystals typically permit phase-matched interactions only between THz frequencies and NIR light near 1550 nm [23], whereas operation at 1030 nm is increasingly desirable because it matches widely available Yb-based ultrafast lasers and enables detection of the upconverted signal with established silicon photodetectors.

In this context, BNA (*N*-benzyl-2-methyl-4-nitroaniline) is particularly promising since it combines phase matching at 1030 nm with a large nonlinear coefficient and a relatively high optical damage threshold, especially when mounted on a thermally conductive holder [23–25]. In this work, we demonstrate broadband, pulsed THz detection based on nonlinear upconversion in BNA using a 1030-nm gating pulse.

A schematic of the upconversion process is shown in Fig. 1(a). We generate and detect sum-frequency generation (SFG) and DFG signals spanning 1 to 7.5 THz with a spectral resolution of 0.2 THz. At 4 THz, the system achieves a detection efficiency of

*Contact author: avish009@uottawa.ca
*Contact author:jean-michel.menard@uottawa.ca

1.98% for SFG photons and a minimum detectable mean THz photon number of better than 0.04 photons per pulse (signal-to-noise ratio, SNR = 1) when averaged over 50,000 pulses. At a higher THz photon flux of ~60 photons per pulse, the system reaches a per-pulse detection probability of 50%. Accounting for both SFG and DFG contributions, we determine the total internal THz-to-NIR photon upconversion efficiency is 76%, with the SFG-only contribution being 45%. Together, these results establish the highest-efficiency room-temperature, free-space, solid-state THz detector demonstrated to date, exceeding previous implementations by an order of magnitude [14,19]. More broadly, they indicate that BNA enables broadband, near single-photon-level THz sensitivity using a room-temperature optical platform.

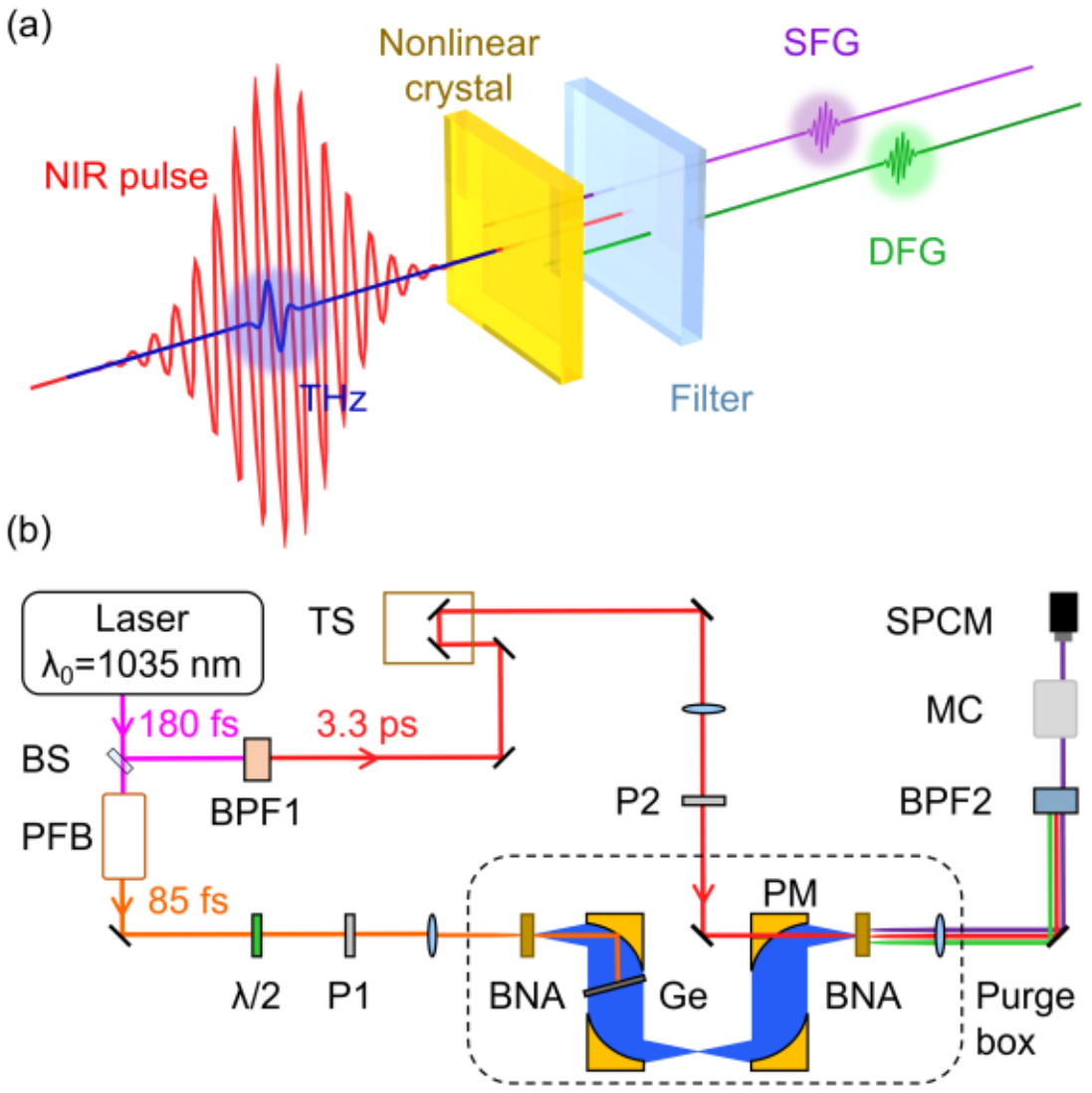

FIG 1. (a) Illustration of the THz detection mechanism based on nonlinear frequency mixing. A near-infrared (NIR) pulse co-propagates with the incoming THz photons inside an organic nonlinear crystal. The $\chi^{(2)}$ interaction generates either SFG or DFG photons. A spectral filter suppresses the residual NIR light, ensuring that only the frequency-converted light is incident on the detector. (b) Schematic of the experimental setup for THz generation and nonlinear detection. BS – beam splitter; PFB – peak field booster [26]; BPF – bandpass filter; BPF1: $\lambda_0$ = 1029.6 nm, $\Delta\lambda$ = 0.6 nm; BPF2: $\lambda_0$ = 1030 nm (1064 nm), $\Delta\lambda$ = 10 nm, angle-tuned by tilting to the SFG (DFG) wavelengths; TS – translation stage; L – lens; P – polarizer; $\lambda/2$ – half-wave plate; BNA – crystal; PM – parabolic mirror; Ge – germanium wafer; MC – monochromator; SPCM – single-photon counting module.

*Contact author: avish009@uottawa.ca
*Contact author:jean-michel.menard@uottawa.ca

## II. METHODS

The experimental setup, illustrated in Fig. 1(b), is driven by an ultrafast laser that delivers 100 µJ, 180 fs pulses centered at 1035 nm and operating at a 50 kHz repetition rate. The beam is divided into two arms using a beam splitter (BS). One arm, carrying approximately 8 nJ, is directed into a peak field booster [26], in which spectral broadening followed by dispersion compensation compresses the pulse. This yields 5 nJ, 85 fs pulses, which are used to achieve broadband THz generation via optical rectification in a 400 µm-thick, <001>-oriented BNA crystal mounted on a sapphire substrate. The second arm is spectrally filtered using a narrowband bandpass filter (BPF1), producing a gating beam with a full-width at half-maximum (FWHM) of 0.6 nm (corresponding to 0.17 THz) and a pulse duration of 3.3 ps. The generated THz pulses and gating beam are brought into spatial and temporal overlap with the gating beam inside a 450 µm-thick, <001>-oriented BNA crystal mounted on a sapphire substrate, enabling frequency upconversion. The resulting SFG and DFG photons are isolated using dedicated BPF2 filters, which can be angle-tuned to transmit different spectral regions. The filtered light is then spectrally resolved using a monochromator and detected by a silicon-based single-photon counting module (SPCM; Excelitas custom module SPCM-CD-34-62-H). The SPCM output is post-gated in a time-tagger (Swabian Time Tagger Ultra) using a 1 ns window synchronized to the laser pulses.

## III. RESULTS AND DISCUSSION

Figure 2 shows the spectra of the SFG and DFG photons measured with the SPCM after filtering out the residual gating beam. The central gap corresponds to the blocked laser region. The overall spectral resolution, determined by the monochromator slit widths and the duration of the gating pulse, is 0.2 THz. The full SFG and DFG spectra are obtained by performing multiple scans over different spectral windows while shifting the passband of BPF2 via small tilt adjustments and subsequently stitching the data together. This approach ensures that every measurement is taken within the flat central region of the filter's transmission band, so that the assembled spectrum reflects only the uniformly transmitted portions of each scan (see Supplemental Material, Sec. S1) [27]. Frequencies below 1 THz cannot be accurately resolved because further tilting of BPF2 shifts the filter roll-off toward 1030 nm, causing the residual gating pulse leakage to increase steeply, eventually dominating the THz-induced upconverted signal and saturating the SPCM. As a result, the response near the lower edge of the recorded spectrum, in the region between 1 and 1.5 THz, must be normalized by the filter transmission spectrum, since the measurement is taken on the sloped roll-off rather than within the flat passband of BPF2. Both the SFG and DFG spectra peak around an absolute frequency of 4 THz and extend up to 7.5 THz, exhibiting nearly symmetrical spectral profiles. The spectral peak and the highest detectable THz frequency are primarily set by the THz generation process, governed by the bandwidth and pulse duration of the NIR pulse after the PFB and by the properties of the BNA generation crystal [23,24]. We observe several dips in the spectrum at 2.2, 3.3, and 5.5 THz, which arise from reported phonon resonances in the BNA crystal [24,29]. The spectra in Fig. 2 are corrected for the detector spectral response and normalized to the SFG peak. For the other experiments discussed in this Letter, we concentrate on the SFG region, where the SPCM exhibits higher photon detection efficiency.

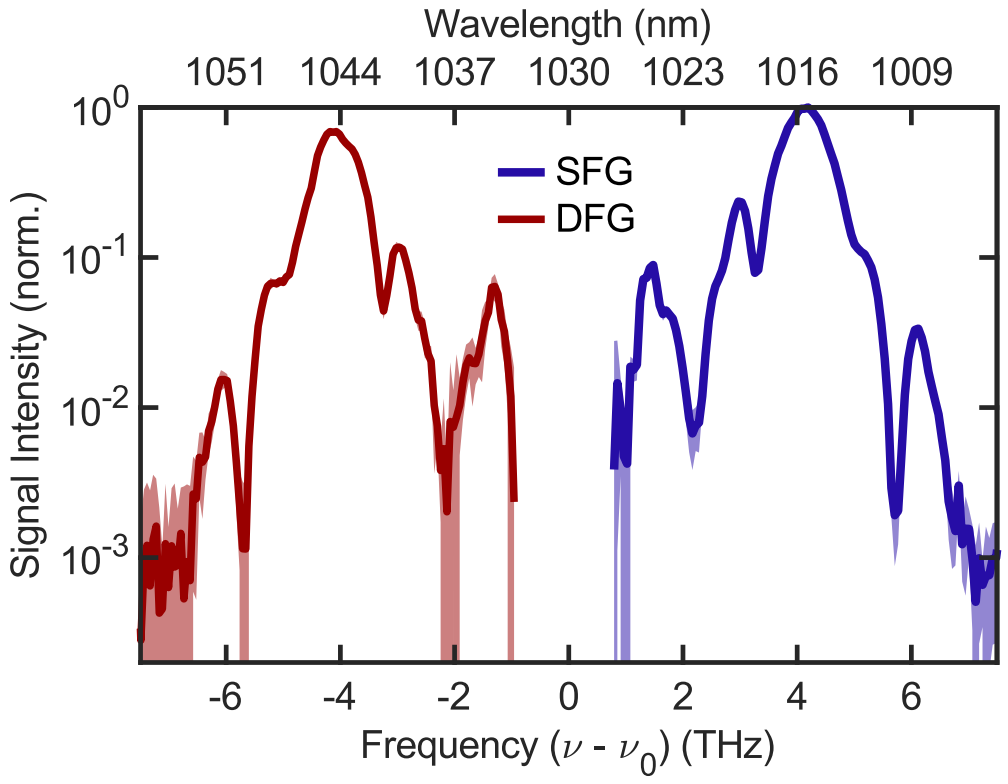

FIG 2. Spectrum of the SFG/DFG signal detected using a 450 µm-thick BNA crystal. The shaded region represents the standard deviation obtained from 10 individual measurements. Central gap corresponds to the blocked spectral region of the gating pulse.

To quantify the detection sensitivity, we measure the detected counts as a function of the incident THz photons per pulse. Each data point in Fig. 3 corresponds to photon counts on the SPCM accumulated over 1 s of gated detection (1 ns electrical gate width), averaged over 180 measurements. The THz pulse energy, and correspondingly the photon number per pulse, are obtained from the calibrated time-domain electro-optic (EO) signal measured using an 85 fs gating pulse and a 200 µm GaP crystal, following the procedure described in previous work [14]. Figure 3(a) shows the THz-induced signal and the background signal, measured by blocking the THz beam, plotted as a function of the incident THz photons per pulse for a peak NIR gating-pulse fluence of 11.4 mJ/cm². The measurements are performed at 1016 nm, corresponding to 4 THz incoming radiation, with the monochromator slits fully open to collect a 1.6 THz spectral bandwidth. The minimum detectable signal corresponds to SFG counts exceeding the background by one standard deviation (SNR = 1), giving a sensitivity of about 0.037 THz photons per pulse within a 1 s averaging time. Figure 3(b) shows the background-subtracted SFG signal versus the incident THz photon number per pulse for three different peak NIR gating-pulse fluences. The curves begin to saturate at SFG count rates exceeding ~0.4 counts per pulse, due to the SPCM's binary response (click or no click). When multiple SFG photons arrive within a single pulse, the detector cannot produce multiple clicks, resulting in a sublinear count rate. The dashed lines show linear fits to the data in the unsaturated regime, and the corresponding slopes yield the full-system detection efficiency for SFG photons, $\eta_{sys}$. For the highest peak NIR gating-pulse fluence of 11.4 mJ/cm², we obtain a $\eta_{sys}$ of 1.98%. At lower fluences, the $\eta_{sys}$ decreases to 1.3% at 5.8 mJ/cm² and 0.24% at 1.2 mJ/cm². An increase of the peak NIR gating-pulse fluence above 22.8 mJ/cm² leads to visible damage (whitening) of the BNA detection crystal. This damage threshold is consistent with reports of increased damage susceptibility for BNA on sapphire substrates in the ~20 mJ/cm² range [30]. The maximum fluence used in our experiments therefore remains a factor of two below this threshold to ensure reproducibility.

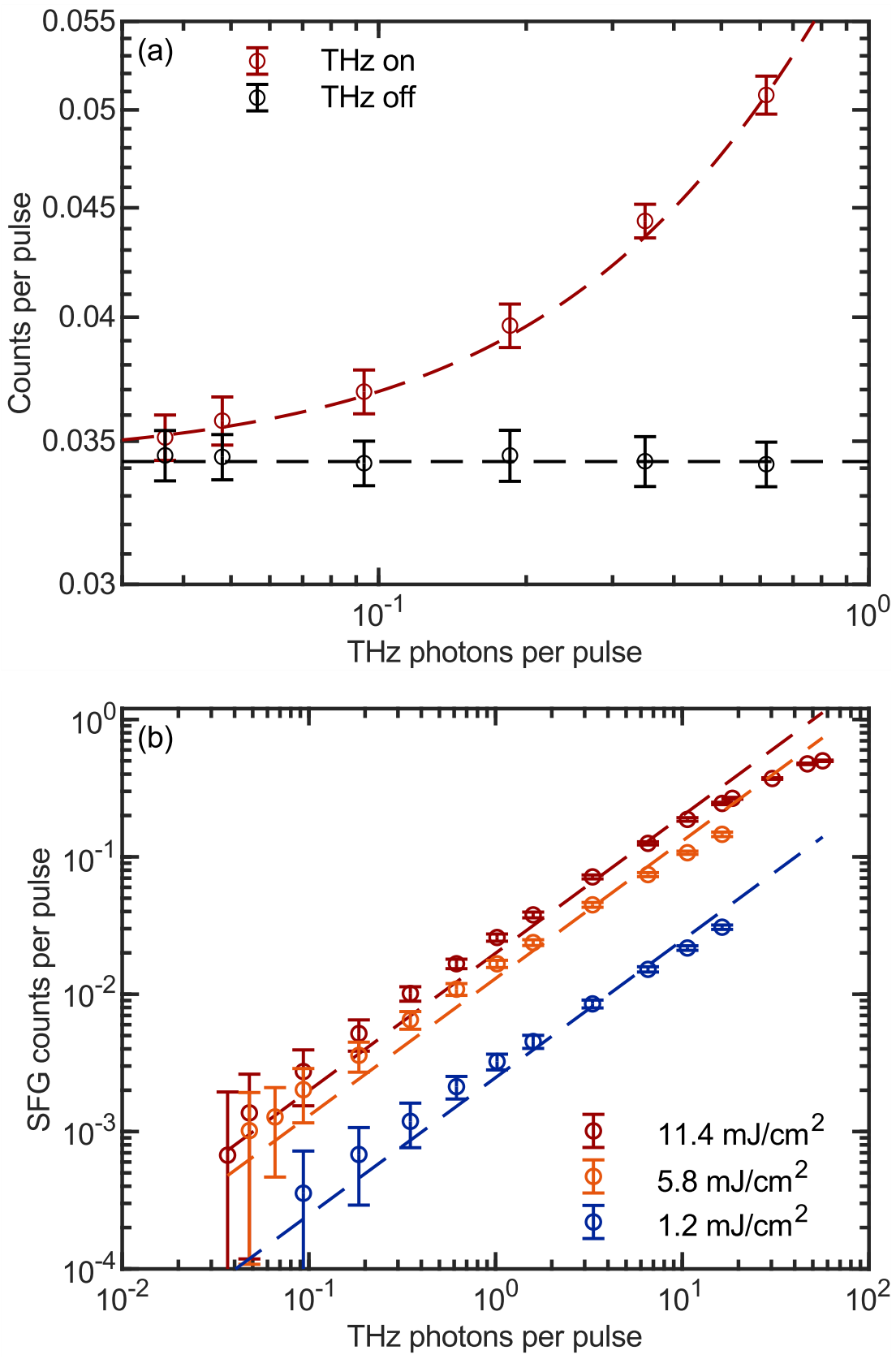

FIG 3. Sensitivity measurements for SFG obtained at 4 THz with the monochromator slits fully open to monitor a 1.6 THz bandwidth. (a) Detected counts per pulse in the presence (red circles) and absence (black circles) of THz radiation, plotted as a function of the incident THz photons per pulse for a peak NIR gating-pulse fluence of 11.4 mJ/cm². The minimum sensitivity level is achieved when the THz signal is one standard deviation above the no-THz background. Each data point corresponds to photon counts accumulated over 1 s of gated detection (1 ns gate width), averaged over 180 measurements. Error bars represent the standard deviation over these 180 measurements. (b) SFG signal, obtained after background subtraction, as a function of incident THz photons per pulse for three different peak NIR gating-pulse fluences. The dashed lines indicate linear fits from which the full-system detection efficiency for SFG photons, $\eta_{sys}$, is extracted.

*Contact author: avish009@uottawa.ca
*Contact author:jean-michel.menard@uottawa.ca

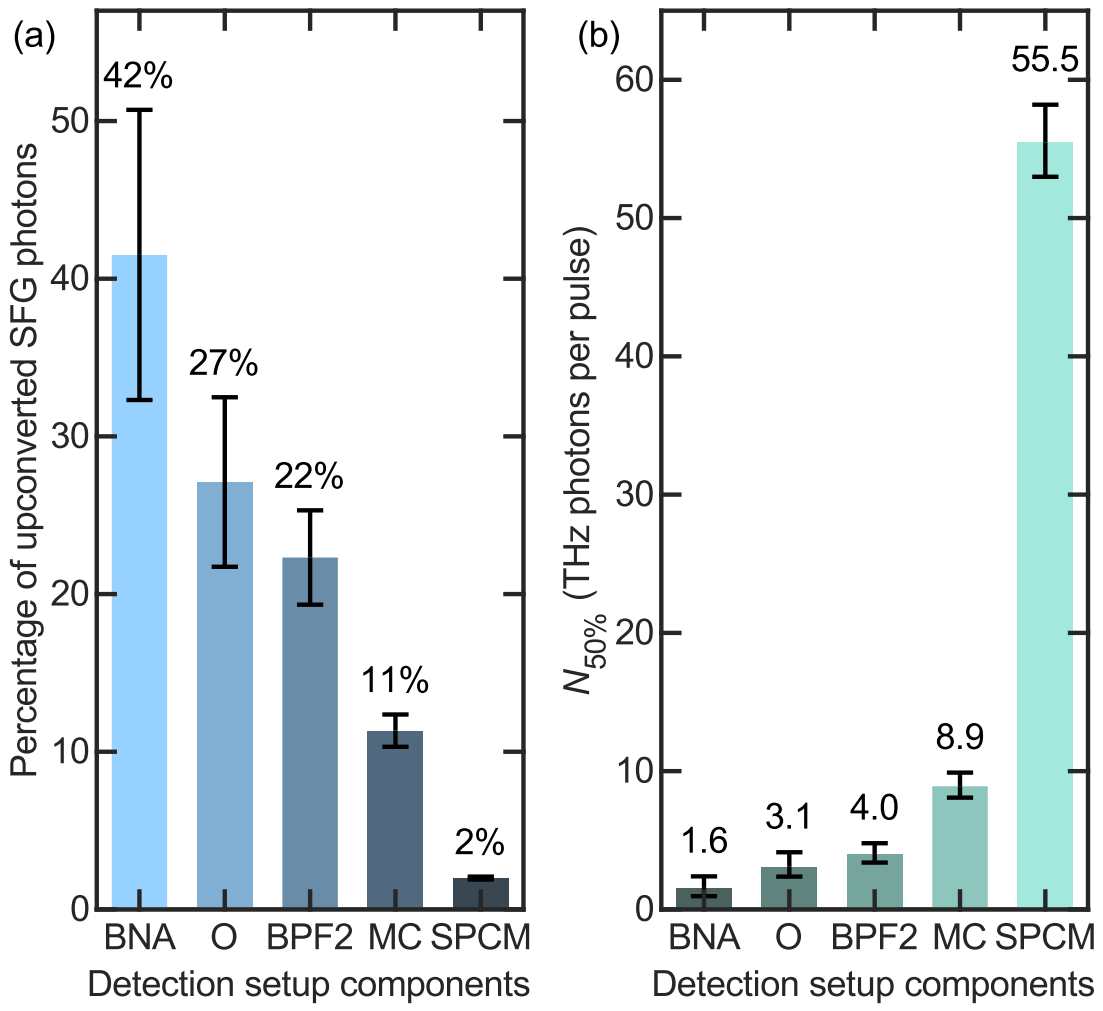

FIG 4. (a) Percentage of upconverted SFG photons present after each component of the detection setup for a peak NIR gating-pulse fluence of 11.4 mJ/cm$^2$. (b) Corresponding number of THz photons per pulse required for a 50% detection probability, $N_{50\%}$, calculated from the percentage of upconverted SFG photons present after each component of the detection setup as the equivalent photon number for an ideal, lossless detector placed immediately after that component. BNA - *N*-benzyl-2-methyl-4-nitroaniline crystal; O - intermediate optics before the bandpass filter; BPF2 - bandpass filter; MC - monochromator; SPCM - single-photon counting module. The percentage of upconverted SFG photons is determined by accounting for the transmission losses of the optical components in the setup. Error bars represent the propagated uncertainty, calculated from the standard deviation of the linear fit used to extract the full-system detection efficiency, $\eta_{sys}$.

*Contact author: avish009@uottawa.ca
*Contact author:jean-michel.menard@uottawa.ca

To pinpoint the origins of loss and guide further optimization, we characterize the optical power loss of each element in the detection path and the photon detection efficiency (PDE) of the SPCM. Using the measured full-system detection efficiency for SFG photons, $\eta_{sys}$, at the SPCM, and accounting for the transmission losses of each of the intervening optical elements and the PDE of the SPCM, we determine the percentage of upconverted SFG photons present just after each component of the detection setup. Equivalently, the percentage of upconverted SFG photons present after each component of the detection setup is given by the ratio of the average number of SFG NIR photons at the output of that element to the number of THz photons incident on the crystal. Figure 4(a) summarizes the percentage of upconverted SFG photons present after each component of the detection setup for a peak NIR gating-pulse fluence of 11.4 mJ/cm². NIR transmission was measured with a Thorlabs S120VC power sensor after each optical element located in the detection beam path including the BNA crystal, the series of optical components preceding the bandpass filter, the bandpass filter BPF2, and the MC, which we also compare to the counts recorded by the SPCM. The SPCM's PDE at 1016 nm is measured to be 17.5 ± 2.2% (see Supplemental Material, Sec. S2 [27]). Error bars indicate the propagated uncertainty, obtained from the 1σ standard deviation of the linear $\eta_{sys}$ fit and the 1σ uncertainties of the measured optical transmission losses and PDE. The percentage of upconverted SFG photons present increases from 1.98±0.09% at the SPCM to 41.5±9.2% immediately after the BNA crystal. After correcting for Fresnel losses at 1016 nm, this corresponds to an in-crystal SFG photon conversion efficiency of 45±10%. Considering the DFG photons, this corresponds to a total in-crystal THz-to-NIR photon conversion efficiency of 76±17%. The primary limitation is the PDE of the SPCM, with additional losses imposed by the monochromator. Hence, further improvements are possible by replacing the monochromator with a lower-loss filtering scheme and potentially shifting the upconverted signal into the visible regime, where the SPCM offers substantially higher PDE approaching 75% at 700 nm.

Based on these results, Fig. 4(b) displays the quantity $N_{50\%}$, defined as the equivalent number of THz photons per pulse incident on the detection crystal that would yield a 50% click probability on an ideal, lossless detector placed at various positions in the setup. This representation makes it possible to compare the intrinsic sensitivity of the system independently of downstream losses. For instance, only 1.6 incident photons per pulse on average at 4 THz are required to reach a 50% click probability on an ideal, lossless detector that is sensitive solely to the SFG photons and positioned immediately after the BNA crystal. The values of $N_{50\%}$ were calculated using a binomial detection model that includes saturation (Supplementary Material, Sec. S3 [27]).

## IV. CONCLUSIONS

We demonstrate a high-sensitivity, room-temperature THz detection scheme based on nonlinear upconversion in the organic crystal BNA using a 1030-nm gating pulse from a Yb-based optical source. The approach supports broadband operation (1–7.5 THz) and converts THz photons near 4 THz into detectable NIR photons with full-system detection efficiency approaching 2%. We reach a 50% per-pulse detection probability at approximately 60 photons per THz pulse. After accounting for external optical losses and detector efficiency, we determine a total in-crystal THz-to-NIR conversion efficiency of 76%. Such an efficient nonlinear conversion shows a pathway toward THz single-photon sensitivity with improved collection and detection efficiency. More broadly, these results establish organic nonlinear crystals as a practical platform for broadband, room-temperature THz photon counting.

## ACKNOWLEDGMENTS

This work was supported by the Natural Sciences and Engineering Research Council of Canada (NSERC) (ALLRP 597331-24, RGPIN-2023-05365) and the National Research Council Canada (NRC) through the Internet of Things: Quantum Sensors Challenge program (QSP-219), the High Throughput and Secure Networks Challenge program (HTSN 254), and the Joint Centre for Extreme Photonics.

## DATA AVAILABILITY

The data that support the findings of this article are not publicly available. The data are available from the authors upon reasonable request.

[1] R. H. Hadfield, Nat. Photonics **3**, 696 (2009).
[2] M. D. Eisaman, J. Fan, A. Migdall, and S. V. Polyakov, Rev. Sci. Instrum. **82**, 071101 (2011).
[3] R. H. Hadfield, J. Leach, F. Fleming, D. J. Paul, C. H. Tan, J. S. Ng, R. K. Henderson, and G. S. Buller, Optica **10**, 1124 (2023).
[4] R. A. Lewis, J. Phys. D: Appl. Phys. **52**, 433001 (2019).
[5] M. Koch, D. M. Mittleman, J. Ornik, and E. Castro-Camus, Nat. Rev. Methods Primers **3**, 48 (2023).
[6] A. G. Markelz and D. M. Mittleman, ACS Photonics **9**, 1117 (2022).
[7] T. Nagatsuma, G. Ducournau, and C. C. Renaud, Nat. Photonics **10**, 371 (2016).
[8] A. Leitenstorfer et al., J. Phys. D: Appl. Phys. **56**, 223001 (2023).
[9] D. F. Santavicca, B. Reulet, B. S. Karasik, S. V. Pereverzev, D. Olaya, M. E. Gershenson, L. Frunzio, and D. E. Prober, Appl. Phys. Lett. **96**, 083505 (2010).
[10] S. Komiyama, O. Astafiev, V. Antonov, T. Kutsuwa, and H. Hirai, Nature **403**, 405 (2000).
[11] M. J. Khan, J. C. Chen, and S. Kaushik, Opt. Lett. **32**, 3248 (2007).
[12] F. Qi, S. Fan, T. Notake, K. Nawata, T. Matsukawa, Y. Takida, and H. Minamide, Opt. Lett. **39**, 1294 (2014).
[13] H. Sakai, K. Kawase, and K. Murate, Opt. Lett. **45**, 3905 (2020).
[14] D. J. Jubgang Fandio, A. Vishnuradhan, E. K. Yalavarthi, W. Cui, N. Couture, A. Gamouras, and J.-M. Ménard, Opt. Lett. **49**, 1556 (2024).
[15] W. Shi, Y. J. Ding, N. Fernelius, and F. K. Hopkins, Appl. Phys. Lett. **88**, 101101 (2006).
[16] M. J. Khan, J. C. Chen, Z.-L. Liau, and S. Kaushik, IEEE J. Sel. Top. Quantum Electron. **17**, 79 (2011).
[17] Q. Fu, P. Liu, W. Li, Y. Yang, F. Qi, W. Li, Y. Wang, and Z. Liu, Opt. Lett. **50**, 6662 (2025).
[18] P. Liu, Q. Fu, K. Zhang, X. Zhang, X. Guo, W. Li, F. Qi, W. Li, and Y. Wu, Opt. Lett. **49**, 7052 (2024).
[19] D. Li, Z. Bai, X. Zuo, Y. Wu, J. Sheng, and H. Wu, Appl. Phys. Rev. **11**, 041420 (2024).
[20] M. Kutas, B. Haase, P. Bickert, F. Riexinger, D. Molter, and G. von Freymann, Sci. Adv. **6**, eaaz8065 (2020).
[21] W. Li, P. Liu, Q. Fu, F. Qi, X. Guo, W. Li, Y. Wang, and Z. Liu, Opt. Laser Technol. **180**, 111443 (2025).
[22] J. Yuan, Q. Guo, X. Zhang, N. Liu, X. Yin, N. Ming, L. Guo, B. Jiao, K. Wang, and S. Fan, Sensors **24**, 6245 (2024).
[23] S. Mansourzadeh, T. Vogel, A. Omar, T. O. Buchmann, E. J. R. Kelleher, P. U. Jepsen, and

*Contact author: avish009@uottawa.ca
*Contact author:jean-michel.menard@uottawa.ca

C. J. Saraceno, Opt. Mater. Express **13**, 3287 (2023).
[24] K. Miyamoto, S. Ohno, M. Fujiwara, H. Minamide, H. Hashimoto, and H. Ito, Opt. Express **17**, 14832 (2009).
[25] W. Cui, A. Vishnuradhan, M. Lippl, E. K. Yalavarthi, A. Gamouras, N. Joly, and J.-M. Ménard, arXiv:2601.11764 [physics.optics].
[26] N. Couture, J. Schlosser, A. Ahmed, M. Wahbeh, G. Best, A. Gamouras, and J.-M. Ménard, Appl. Opt. **62**, 4097 (2023).
[27] See Supplemental Material at [URL will be inserted by publisher] for details of the construction of the full SFG and DFG spectra (Sec. S1), photon-detection efficiency (PDE) measurement (Sec. S2), and the saturated binomial detection model (Sec. S3), which includes Ref. [28].
[28] ISS, “Single Photon Counting Modules (SPADs),” https://iss.com/fluorescence/modules-components/light-detectors/spads-single-photon-counting-modules#specifications (accessed Mar. 25, 2026).
[29] S. Mansourzadeh, T. Vogel, A. Omar, M. Shalaby, M. Cinchetti, and C. J. Saraceno, APL Photonics **8**, 011301 (2023).
[30] K. Brekhov, S. Colar, E. Chiglintsev, and A. Chernov, Opt. Express **33**, 30505 (2025).

*Contact author: avish009@uottawa.ca
*Contact author:jean-michel.menard@uottawa.ca